\documentclass[a4paper,11pt]{article}

\usepackage{pos}
\usepackage{subcaption}
\usepackage{wrapfig}

\title{Mapping the ice stratigraphy in IceCube\newline using camera deployment footage
}

\ShortTitle{IceCube camera logging}

\author{The IceCube Collaboration \\{\normalsize \normalfont(a complete list of authors can be found at the end of the proceedings)}\\}

\emailAdd{anna.eimer@fau.de}
\emailAdd{martin.rongen@fau.de}

\abstract{

The IceCube Neutrino Observatory is a cubic-kilometer Cherenkov array deployed in the deep, glacial ice at the geographic South Pole. An important feature of the instrumented ice are undulations of layers of constant optical properties over the footprint of the detector. During detector construction, these layers were mapped using stratigraphy measurements obtained from a stand-alone laser dust logger. While this system is very precise, its cost does not scale to the instrumented volume envisioned for the proposed IceCube-Gen2 Observatory. Here, we explore the possibility of obtaining equivalent stratigraphy data from camera footage recorded during the deployment of IceCube more than a decade ago. If successful, this could be an alternative technique to be considered for IceCube-Gen2.

\vspace{4mm}

{\bfseries Corresponding authors:}
Anna Eimer$^{1*}$, 
Martin Rongen$^{1}$\\
{$^{1}$ \itshape Erlangen Centre for Astroparticle Physics, Friedrich-Alexander-Universität Erlangen-Nürnberg}\\
[4mm]
$^*$ Presenter
}

\ConferenceLogo{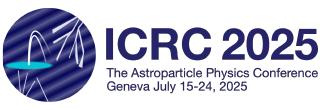}

\FullConference{39th International Cosmic Ray Conference (ICRC2025)\\
 15–24 July 2025\\
Geneva, Switzerland\\}

\begin{document}

\maketitle

\section{The ice stratigraphy and layer undulations}\label{sec2}

The IceCube Neutrino Observatory \cite{detector_paper} is a Cherenkov telescope instrumenting a cubic kilometer of deep, glacial ice at the geographic South Pole. As the Cherenkov light, emitted from charged relativistic particles, propagates through the ice it is absorbed or scattered by impurities. Therefore, understanding the optical properties of the ice is essential for accurate event reconstruction.

The ice making up the detector today has built up from snow precipitation over roughly 100'000 years \cite{journal_of_glaciology_2013}. As the falling snow carried impurities from the atmosphere, the ice exhibits a stratigraphy of impurity concentrations, and thus varying absorption and scattering lengths, reflecting the climatic conditions at the time of deposition.

\begin{figure}[h]
\centering
\includegraphics[width=\linewidth]{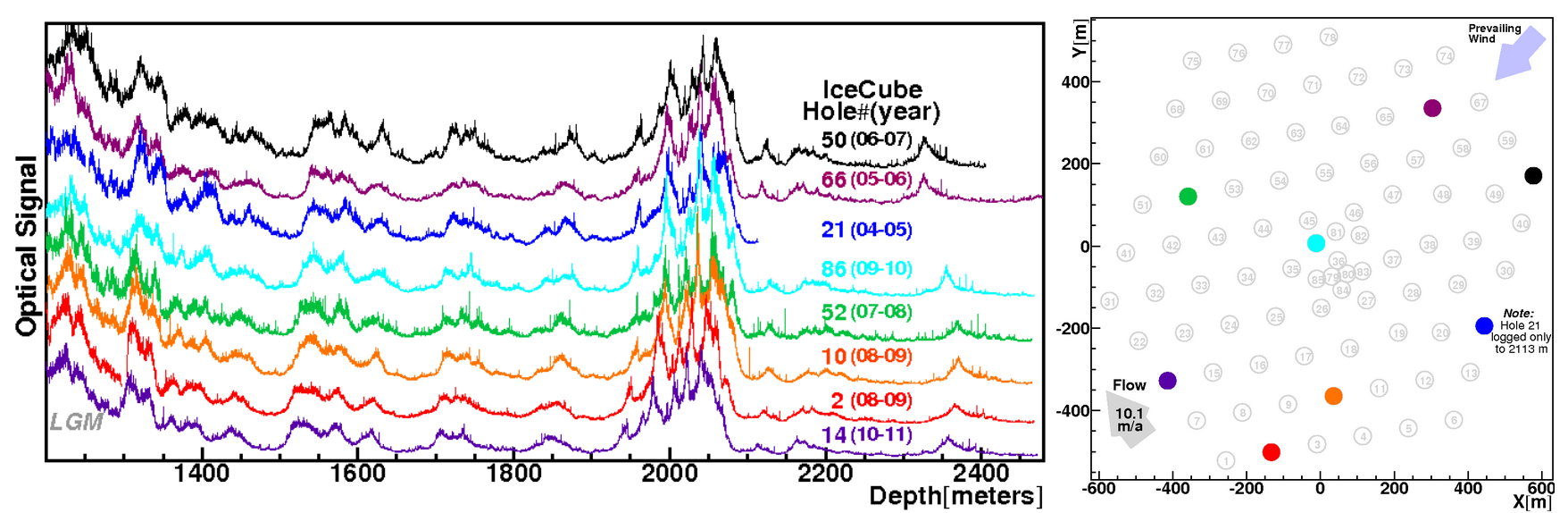}
\caption{Stratigraphies as obtained from the original dust logger. The strength of the optical return signal (left) is plotted as a function of depth for the eight drill holes indicated on the right. Adapted from \cite{journal_of_glaciology_2013}}
\label{oldlogger}
\end{figure}

In IceCube, the scattering stratigraphy was originally mapped with sub-millimeter resolution using a so-called re-usable laser dust logger \cite{journal_of_glaciology_2013}. This device was lowered into eight IceCube drill holes prior to the deployment of the instrumentation cables. During its round trip, it continuously shone laser light horizontally into the ice. A small fraction of this light was back-scattered on impurities and reached a 1" photomultiplier at the bottom of the logger. The resulting stratigraphies are shown in Figure \ref{oldlogger}. 
Strikingly, the eight dust logger stratigraphies show characteristic features at different absolute depths. This is the result of ice isochrones, which are ice layers deposited at the same time and thus featuring the same optical properties, conforming to the shape of the underlying bedrock at depth. The three-dimensional shape of these so-called ice layer undulations was interpolated from the dust logger data \cite{journal_of_glaciology_2013} and only recently updated using fits to LED calibration data \cite{LEDtilt:ICRC2023}.

The mapping of ice layer undulations will be of vital importance for the envisioned IceCube-Gen2 array \cite{Gen2paper}. Here, the larger lateral spacing between strings will likely require the logging of every drill hole. As the holes start refreezing immediately after drilling, the deployment of the re-usable dust logger necessitates drilling larger hole diameters. This adds significant fuel costs and logistical burden to the project. Consequently, efforts are underway to establish logging methods that may be performed during the deployment of the instrumentation strings.

Some of these methods will be tested during the deployment of the IceCube Upgrade \cite{UpgradeProceedings}, during this austral summer season. This, for example, includes a co-deployed laser dust logger \cite{LOMlogger:2025icrc} which uses the existing sensor modules as receivers and only adds a dedicated light source.

We here present a feasibility study to investigate whether camera video footage can also provide useful stratigraphy information. For this study we analyze historic camera footage obtained during the final deployment season of IceCube in 2010/11.
A similar attempt has also already been made using fixed-focus cameras, designed for the IceCube Upgrade \cite{UpgradeCamera}, deployed in the SpiceCore drill hole  \cite{SpiceCoreCamera}. While conceptually successful, the spatial resolution in this study was limited by the required 3\,s integration time. 

\section{The Sweden Camera}

\begin{figure}
\centering
\begin{minipage}{.45\textwidth}
  \centering
    \includegraphics[width=\linewidth]{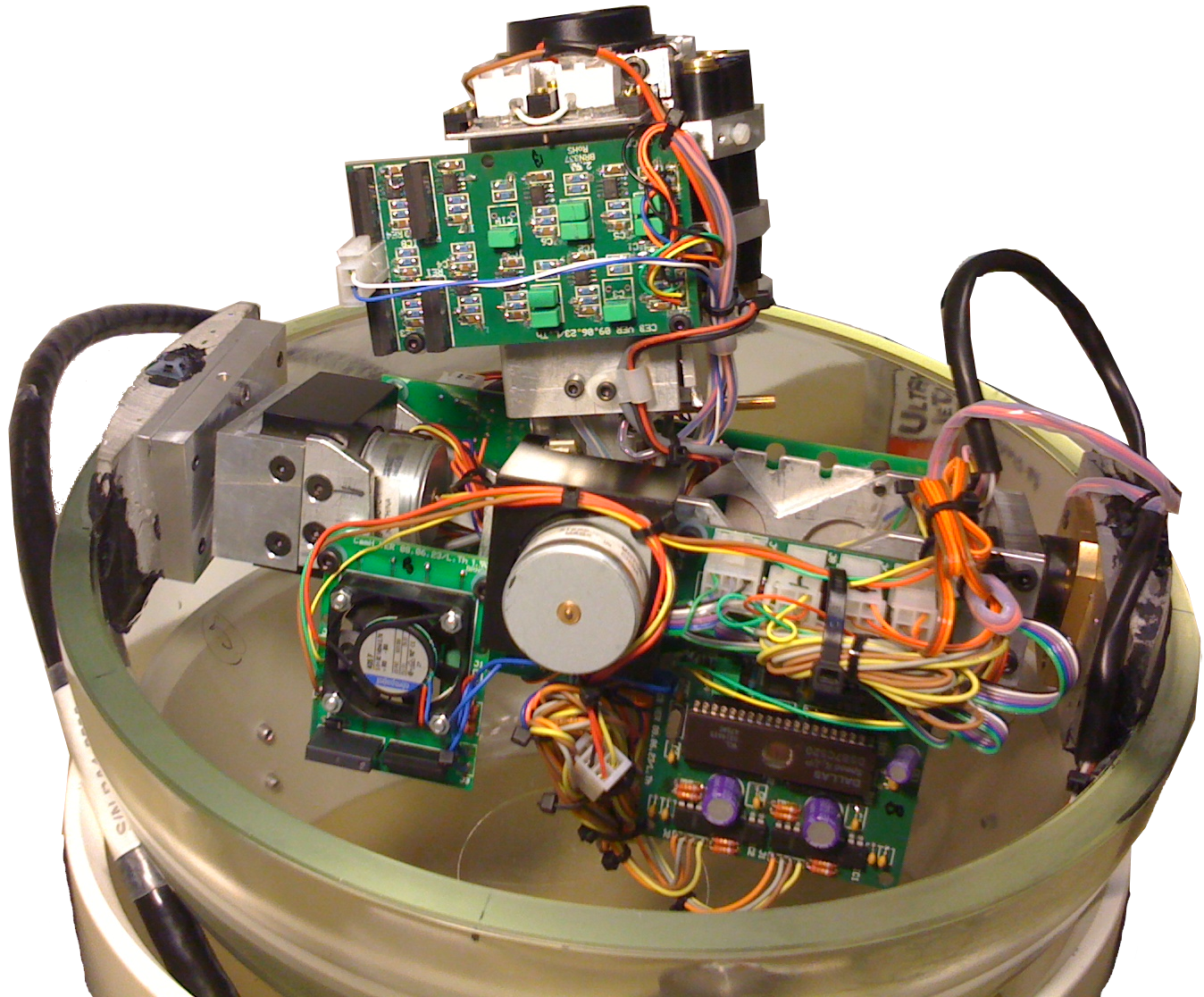}
    \caption{One of the Sweden Cameras during assembly. The actual camera, with its laser light sources to the side, is seen pointing up.}
  \label{fig:hardware}
\end{minipage}%
\hfill
\begin{minipage}{.5\textwidth}
  \centering
    \includegraphics[width=\linewidth]{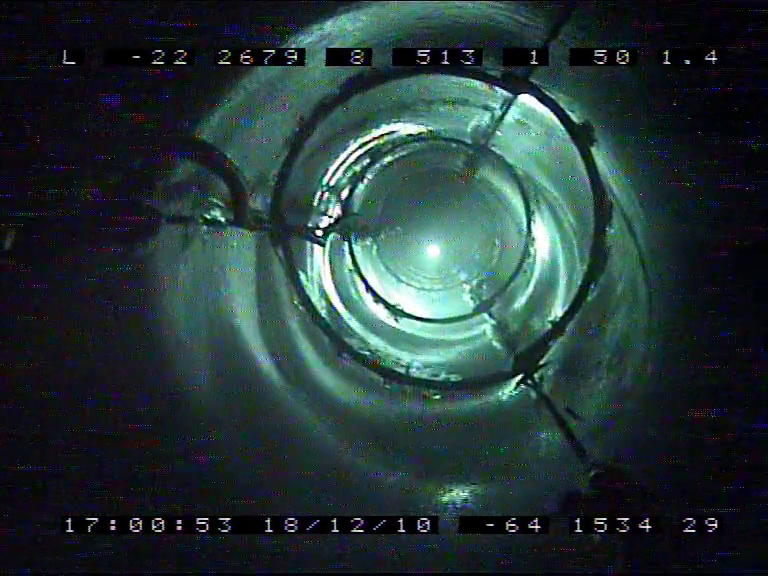}
    \caption{Video still frame of the lower camera looking up the drill hole towards the upper camera which is emitting light.}
  \label{fig:water}
\end{minipage}
\end{figure}

\begin{figure}
\centering
\includegraphics[width=\linewidth]{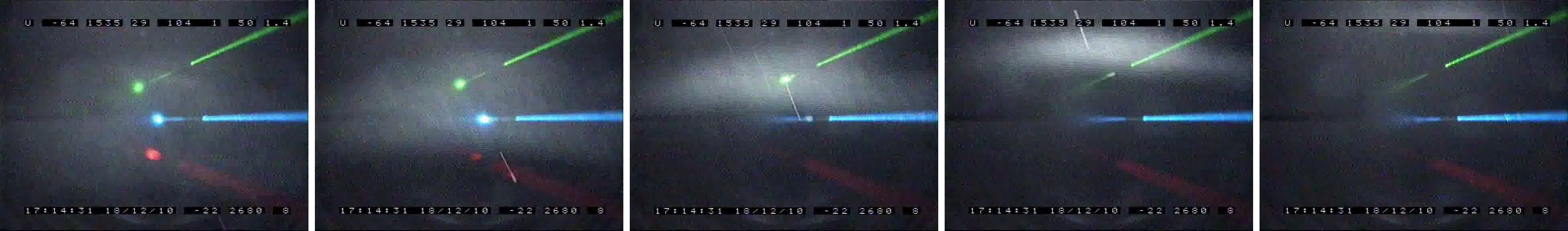}
\caption{Every second frame from a 0.24\,s video sequence of the lasers sweeping over a strong and localized feature at a depth of 2145\,m.}
\label{sequence}
\end{figure}

This study uses a camera system designed at Stockholm University, known as the Sweden Camera, which was deployed on String 80 during the last construction season. Its primary goals were to observe the freeze-in process and to assess the optical characteristics of the drill hole. This system comprises two video cameras encased in separate glass pressure spheres (see Figure \ref{fig:hardware}), positioned 5.8 meters apart. The cameras are capable of rotating to view multiple directions, albeit with certain mechanical restrictions. Both cameras are outfitted with four white LED lights and three lasers in red, blue, and green. The heating, illumination, and movement of the cameras can be remotely controlled via a dedicated surface system. Each camera transmits an analog video feed to the surface, where only the signal from one camera can be digitized at any time. The digitized video also contains an overlay displaying the camera orientation and imaging settings at any given time.

\section{Camera deployment footage}

\begin{figure}
\centering
\includegraphics[width=0.75\linewidth]{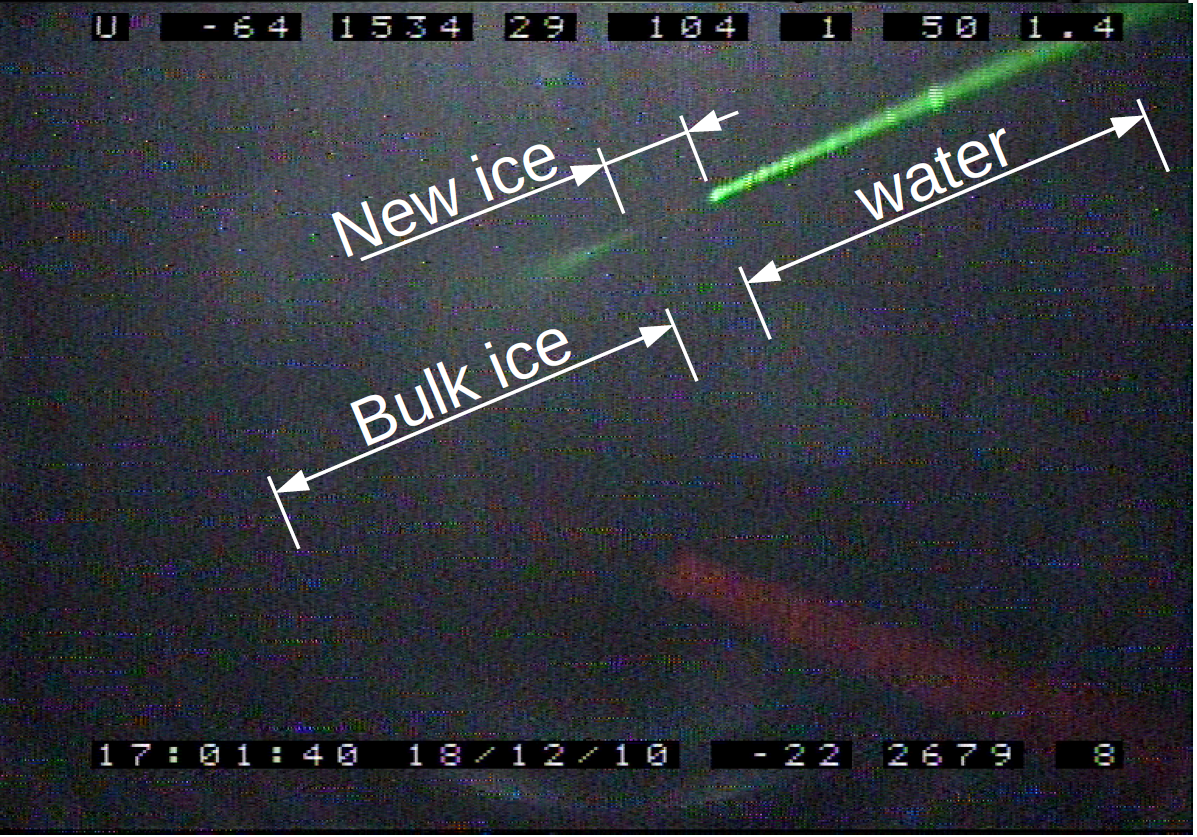}
\caption{Example individual video frame. The laser light originates at the upper right corner of the image. It is strongly scattered in the water column of the drill hole into which the camera is being deployed. Very little scattering intensity is seen in the already refrozen outer layer. The scattering in the bulk glacial ice is used to build the ice stratigraphy.}
\label{fig:individual}
\end{figure}

The Sweden Camera system was operated during the entire assembly and drop of the string, resulting in over nine hours of footage. We here focus on the longest continuous drop segment from a depth of 1733\,m to 2256\,m, spanning 30 minutes. The video was digitized at a resolution of 768\,px\,$\cdot$\,576\,px and with 25 frames per second. So given an average drop speed of approximately 18 meters per minute each frame covers around 12\,mm in depth.

During the drop the upper camera was mostly operated at constant settings, with a focal length of 104\,mm (focal ratio of 1.4) and shutter speed of 1/50\,s. While these settings are constant, there appears to be some undocumented automatic variable gain at play. This is evident by the video initially saturating as bright light sources are turned on, with the video brightness then slowly adjusting to the new condition. 
As the equatorial harness of the pressure housing prevents the camera from looking out horizontally, it was set up looking down at an angle of $26.4^{\circ}$. As a result, planar ice features are seen slightly from the top and the intersecting laser spot wanders over the feature as the camera descends. The laser beam is pointing towards the image center, but the exact pointing of the laser is unknown. The lower camera was mostly used to view up and down the water column (see Figure \ref{fig:water}) and is not used in this study. Overall, the video material is rather inconsistent, with operators manually switching between cameras, enabling or disabling light sources or changing the focus length. 
Out of the total of 44062 frames making up this drop segment, 10600 were manually removed from the analysis for inconsistent acquisition settings. In the frames analyzed, the green laser is always enabled and brightly visible. The white LEDs are mostly on, with only two short gaps covering a total of $\sim$30\,m.

\begin{figure}
\centering

\tabskip=10pt
\valign{#\cr
    \vspace{0.5cm}
  \hbox{%
    \begin{subfigure}{.45\textwidth}
    \centering
    \includegraphics[width=\textwidth]{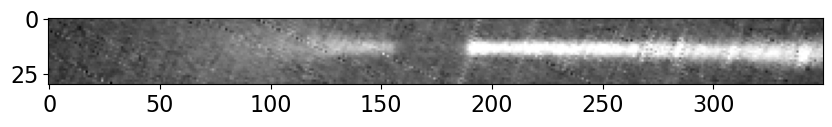}
    \caption{Grey-scale image of the green channel in the region around the green laser beam. Note the region of new ice with minimal scattering around pixel number 175. The bulk ice is to the left and the water column is to the right.}
    \label{beam:mask}
    \end{subfigure}%
  }\vfill
  \hbox{%
    \begin{subfigure}{.45\textwidth}
    \centering
\includegraphics[width=\textwidth]{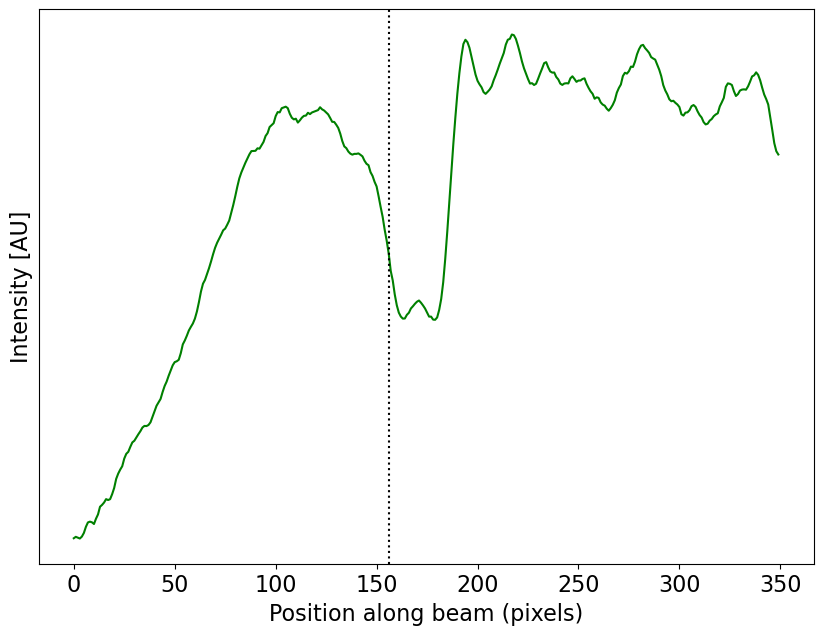}
    \caption{Intensity profile along the beam obtained by integrating Figure \ref{beam:mask} over the y-axis. The fitted transition into the bulk ice is denoted as a dotted line.}
    \label{beam:profile}
    \end{subfigure}%
  }\cr
 \noalign{\hfill}
  \hbox{%
    \begin{subfigure}[b]{.5\textwidth}
    \centering
   \includegraphics[width=\textwidth]{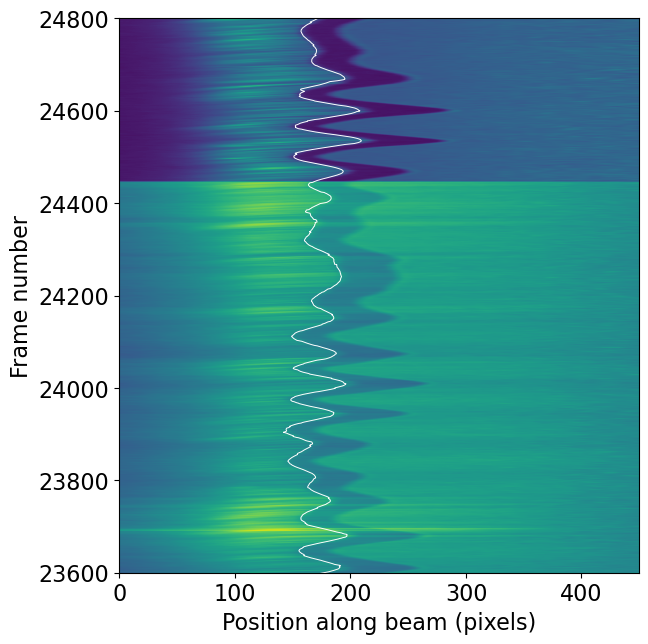}
    \caption{Synthesized stratigraphy image obtained by stacking intensity profiles from 1200 video frames ($\sim$14.5\,m). The fitted transition into the bulk ice is overlaid as a white line. The bulk ice stratigraphy shown in Figure \ref{green:stratigraphy} is obtained as the average intensity at a set distance from the bulk ice transition. For frame numbers larger than 24450 the white LEDs were off, while for smaller frame numbers the white LEDs were on, resulting in a loss of contrast.}
    \label{intensity:stack}
    \end{subfigure}%
  }\cr
}
\caption{Overview of the video processing performed to obtain the stratigraphy.}
\label{}
\end{figure}

\section{Frame by frame image processing}

The image analysis is performed frame by frame. For an example frame see Figure \ref{fig:individual}. The laser light originates from the upper right corner of the image. It is strongly scattered in the water column. Very little scattering intensity is seen in the already refrozen outer layer of the drill hole. The intensity of scattering in the bulk glacial ice needs to be quantified to build the stratigraphy.

For each frame the green channel is extracted, cropped to a small region around the green laser beam and rotated so that the beam is nearly horizontal (see Figure \ref{beam:mask}). The intensity profile along the beam axis is then obtained by integrating over the direction orthogonal to the beam and applying a moving averaging filter with a width of 11 pixels along the beam axis. An example intensity profile is seen in Figure \ref{beam:profile}. In these profiles, the new ice is identified as the broad minimum between the brighter water and bulk ice regions. The transition from the newly refrozen ice into the glacial bulk ice is then identified as the coordinate of the maximum gradient to the left of the new ice. 

The intensity profiles from all frames are stacked to form a synthesized stratigraphy image. A small section of this is shown in Figure \ref{intensity:stack}. The newly refrozen ice is seen as a dark band. As the camera descends, it bumps against the walls of the drill hole, resulting in a quasi-periodic shift of the position and depth of the new ice in the image coordinates. The identified transition into the bulk glacial ice (after smoothing over 21 frames) is marked as the white line in Figure \ref{intensity:stack}.
The intensity of the water region (right of the new ice) can be seen to vary smoothly, while the intensity of the glacial bulk ice (left of the new ice) shows more discrete variations as a function of frame number. The banding visible in the water region is mostly a result of diffusely back-scattered white light, some of which originates from the bulk ice, and is not observed in sections where the white LED light is disabled.

In addition to the green laser, we also analyzed the overall image brightness resulting from the white LEDs. For this purpose, we defined an analysis region parallel to the green laser and repeated the analysis as before.

\section{Resulting stratigraphy}

From the synthesized stratigraphy image a one-dimensional stratigraphy can be obtained by evaluating the average brightness for a given horizontal pixel range starting a given margin away from the bulk ice transition. 

\begin{figure}[h]
\centering
\includegraphics[width=\linewidth]{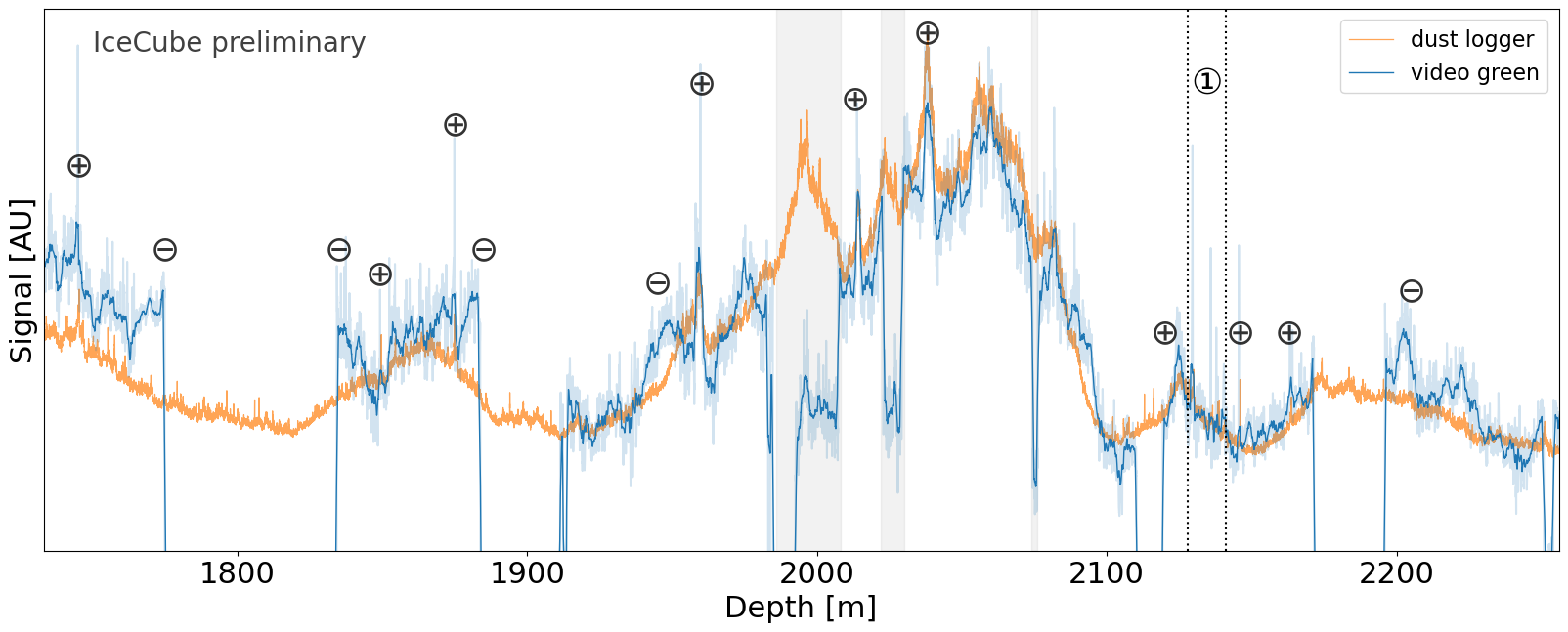}
\caption{Overall stratigraphy obtained from the analysis of the green laser data. The pastel-blue line has been smoothed over seven consecutive frames, while the solid-blue line has been smoothed over 100 frames. The white LED light was turned off in regions with a gray shaded background. No useful video data was available in the regions without a blue curve. Data from the original dust logger in the nearby string 86 is shown for comparison. The stratigraphies have been aligned based on pressure sensor depth readings. The annotated features are discussed in the text.}
\label{green:stratigraphy}
\end{figure}

\begin{figure}[h]
\centering
\includegraphics[width=\linewidth]{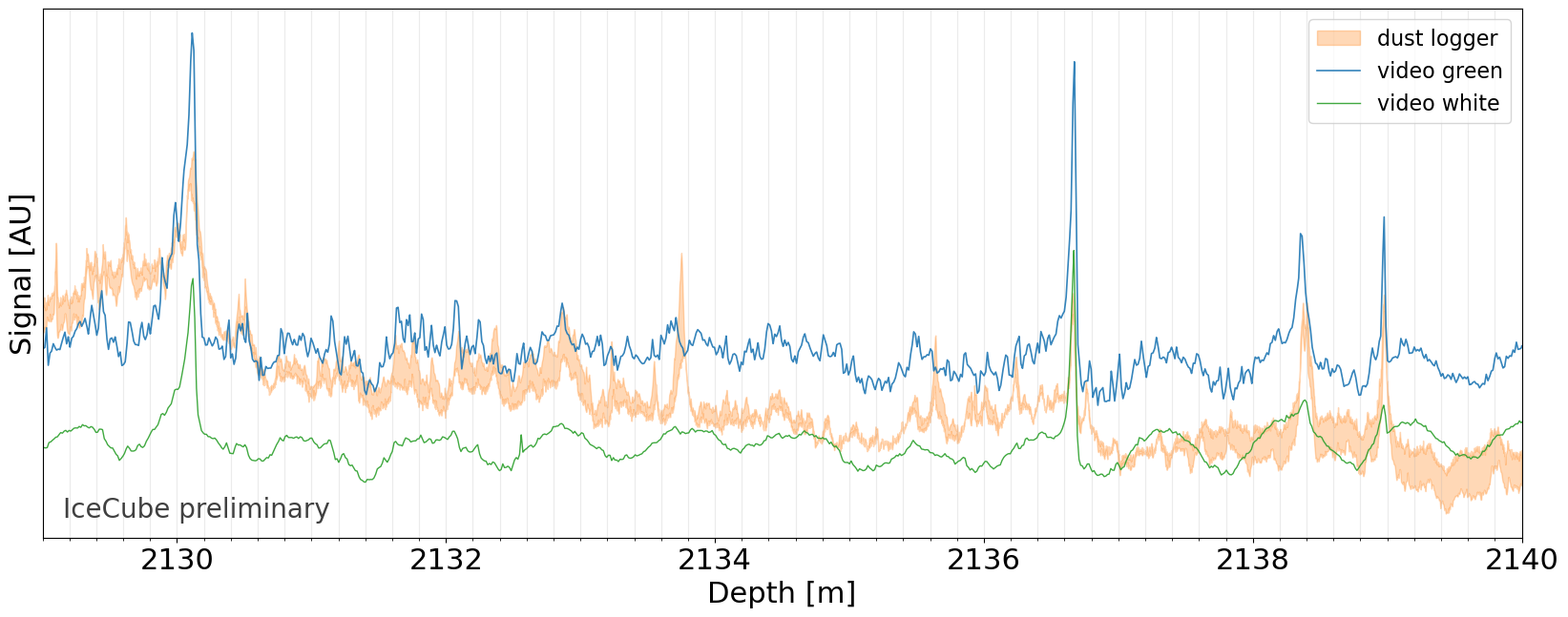}
\caption{11\,m sub-region at a depth around 2134\,m (indicated as \raisebox{.5pt}{\textcircled{\raisebox{-.9pt} {1}}} in Figure \ref{green:stratigraphy}). While the stratigraphy derived from the video footage shows significant artifacts, the strongest features can be clearly identified and are even retained in the analysis of the light from the white LEDs. The width of the orange curve indicates the discrepancy between logging hole 86 during descent and ascent of the old dust logger.}
\label{detail:reggion}
\end{figure}

For the stratigraphies shown in Figures \ref{green:stratigraphy} and \ref{detail:reggion} the margin was chosen as 10 pixels, to sample the closest bulk ice possible while still accounting for the smoothing applied to the intensity profiles, and the averaging range was chosen to be 20 pixels. A larger pixel range would include larger statistics but also induce an increased depth smearing due to the inclined viewing angle of the camera. 
The frame numbers have been converted to absolute depth via pressure readings from a pressure sensor \cite{detector_paper}, taking into account the distance between the pressure sensor and the camera as well as timing offsets between the two readout systems.\newline

To be able to compare to previous measurements, the dust logger data from the closest logged hole (string 86) is shown in the same Figures. As the camera stratigraphy has comparatively little dynamic range, the dust logger signal was square-rooted and scaled to visually match the amplitude of the variations observed in the camera analysis.
A quantitative comparison between the measurements is still challenging because the dust logger stratigraphy can not be treated as ground truth, as it was not obtained from the same drill hole.

In addition, repeated logs of the same hole also show differences due to different azimuthal orientations of the laser and the previously observed optical anisotropy effect of the ice \cite{logger:anisotropy}. The width of the orange curve in Figure \ref{detail:reggion} exemplifies these variations for an ascending and a descending log of hole 86. \newline

\noindent Given Figure \ref{green:stratigraphy}, we can still draw the following conclusions:
\begin{itemize}
    \setlength\itemsep{-0.1em}
    \item The overall shape of the expected stratigraphy, with the strongly scattering "dust layer" around between 1950\,m and 2150\,m and cleaner surrounding ice is generally reproduced.
    \item Very localized features (some of which are marked with \raisebox{.5pt}{\textcircled{\raisebox{-.9pt} {+}}}) are very well matched both in shape and depth localization. An example sequence of images taken by the camera passing by such a local feature is shown in Figure \ref{sequence}.
    \item On intermediate scales the agreement between the dust logger and the camera stratigraphy is often poor. In many cases, such as marked with \raisebox{.5pt}{\textcircled{\raisebox{-.9pt} {-}}}, the camera stratigraphy shows variations that are not expected.  There is some evidence that some of these artifacts stem from the camera moving around in the water column. As the thickness of the water column changes, equivalent features in the ice may appear brighter or dimmer simply because they move closer or further away. Some artifacts may also be caused by the dynamic gain adjustment of the camera.
\end{itemize}


\newpage

As argued in \cite{LOMlogger:2025icrc}, the primary feature size needed to model the ice layer undulations is on the decimeter scale. To be able to gauge the performance of the camera study on this scale, we examine a small subregion with many such decimeter features as shown in Figure \ref{detail:reggion}. The four most prominent dust logger features at 2130\,m, 2136.6\,m, 2138.4\,m and 2139\,m are all well reproduced in the camera stratigraphy. The two strongest features are even clearly evident in the white light stratigraphy.

\section{Conclusion}

The measurement of ice scattering stratigraphies at multiple locations within ice Cherenkov telescopes is important to model the spatial undulation of ice isochrons. We have demonstrated the feasibility of obtaining a sub-decimeter-resolution stratigraphy using high frame-rate video data a camera being deployed into a water-filled drill hole. The available camera footage from the original deployment of the IceCube Neutrino Observatory in 2010 was not ideal for this purpose, with many gaps in usable data and with white LED light reducing the contrast of the primarily investigated laser beams. In addition, unknown gain variations and uncontrolled movements of the camera within the water column likely introduced artifacts into the stratigraphy. 
Nevertheless, decimeter ice features, as most relevant for the modeling of layer undulations, could be reliably identified and matched to the stratigraphy obtained through traditional millimeter-resolution laser dust logging. 

The camera-based and other proposed logging methods will be further developed and tested during the upcoming deployment of the IceCube Upgrade, this austral summer season. With multiple feasible options, the choice of method for stratigraphy logging in the planned IceCube-Gen2 array will likely be driven by practical considerations.

\bibliographystyle{ICRC}
\bibliography{references}

\clearpage

\section*{Full Author List: IceCube Collaboration}

\scriptsize
\noindent
R. Abbasi$^{16}$,
M. Ackermann$^{63}$,
J. Adams$^{17}$,
S. K. Agarwalla$^{39,\: {\rm a}}$,
J. A. Aguilar$^{10}$,
M. Ahlers$^{21}$,
J.M. Alameddine$^{22}$,
S. Ali$^{35}$,
N. M. Amin$^{43}$,
K. Andeen$^{41}$
% ...
\\
\\
% ...
$^{10}$ Universit{\'e} Libre de Bruxelles, Science Faculty CP230, B-1050 Brussels, Belgium \\
% ...
$^{16}$ Department of Physics, Loyola University Chicago, Chicago, IL 60660, USA \\
$^{17}$ Dept. of Physics and Astronomy, University of Canterbury, Private Bag 4800, Christchurch, New Zealand \\
% ...
$^{21}$ Niels Bohr Institute, University of Copenhagen, DK-2100 Copenhagen, Denmark \\
$^{22}$ Dept. of Physics, TU Dortmund University, D-44221 Dortmund, Germany \\
% ...
$^{35}$ Dept. of Physics and Astronomy, University of Kansas, Lawrence, KS 66045, USA \\
% ...
$^{39}$ Dept. of Physics and Wisconsin IceCube Particle Astrophysics Center, University of Wisconsin{\textemdash}Madison, Madison, WI 53706, USA \\
% ...
$^{41}$ Department of Physics, Marquette University, Milwaukee, WI 53201, USA \\
% ...
$^{43}$ Bartol Research Institute and Dept. of Physics and Astronomy, University of Delaware, Newark, DE 19716, USA \\
% ...
$^{63}$ Deutsches Elektronen-Synchrotron DESY, Platanenallee 6, D-15738 Zeuthen, Germany \\
$^{\rm a}$ also at Institute of Physics, Sachivalaya Marg, Sainik School Post, Bhubaneswar 751005, India \\
% ...

\subsection*{Acknowledgments}

\noindent
The authors gratefully acknowledge the support from the following agencies and institutions:
USA {\textendash} U.S. National Science Foundation-Office of Polar Programs,
U.S. National Science Foundation-Physics Division,
U.S. National Science Foundation-EPSCoR,
U.S. National Science Foundation-Office of Advanced Cyberinfrastructure,
Wisconsin Alumni Research Foundation,
Center for High Throughput Computing (CHTC) at the University of Wisconsin{\textendash}Madison,
Open Science Grid (OSG),
Partnership to Advance Throughput Computing (PATh),
Advanced Cyberinfrastructure Coordination Ecosystem: Services {\&} Support (ACCESS),
Frontera and Ranch computing project at the Texas Advanced Computing Center,
U.S. Department of Energy-National Energy Research Scientific Computing Center,
Particle astrophysics research computing center at the University of Maryland,
Institute for Cyber-Enabled Research at Michigan State University,
Astroparticle physics computational facility at Marquette University,
NVIDIA Corporation,
and Google Cloud Platform;
Belgium {\textendash} Funds for Scientific Research (FRS-FNRS and FWO),
FWO Odysseus and Big Science programmes,
and Belgian Federal Science Policy Office (Belspo);
Germany {\textendash} Bundesministerium f{\"u}r Forschung, Technologie und Raumfahrt (BMFTR),
Deutsche Forschungsgemeinschaft (DFG),
Helmholtz Alliance for Astroparticle Physics (HAP),
Initiative and Networking Fund of the Helmholtz Association,
Deutsches Elektronen Synchrotron (DESY),
and High Performance Computing cluster of the RWTH Aachen;
Sweden {\textendash} Swedish Research Council,
Swedish Polar Research Secretariat,
Swedish National Infrastructure for Computing (SNIC),
and Knut and Alice Wallenberg Foundation;
European Union {\textendash} EGI Advanced Computing for research;
Australia {\textendash} Australian Research Council;
Canada {\textendash} Natural Sciences and Engineering Research Council of Canada,
Calcul Qu{\'e}bec, Compute Ontario, Canada Foundation for Innovation, WestGrid, and Digital Research Alliance of Canada;
Denmark {\textendash} Villum Fonden, Carlsberg Foundation, and European Commission;
New Zealand {\textendash} Marsden Fund;
Japan {\textendash} Japan Society for Promotion of Science (JSPS)
and Institute for Global Prominent Research (IGPR) of Chiba University;
Korea {\textendash} National Research Foundation of Korea (NRF);
Switzerland {\textendash} Swiss National Science Foundation (SNSF).

\end{document}